\documentclass[pre,floatfix,twocolumn,showpacs,preprintnumbers,amsmath,amssymb,nofootinbib]{revtex4}

\usepackage[dvips]{graphicx}
\begin{document}
\preprint{Preprint \today}
\title{Resonant control of stochastic spatio-temporal dynamics in a tunnel diode by
multiple time delayed feedback} 

\author{Niels Majer}
%\email{majer@physik.tu-berlin.de}
\author{Eckehard Sch{\"o}ll}
\email{schoell@physik.tu-berlin.de}

\affiliation{Institut f{\"u}r Theoretische Physik, Technische Universit{\"a}t 
Berlin, D-10623 Berlin, Germany}

\begin{abstract}
We study the control of noise-induced spatio-temporal current density patterns 
in a semiconductor nanostructure (double barrier resonant tunnelling diode)
by multiple time-delayed feedback. 
We find much more pronounced resonant features of noise-induced oscillations
compared to single time feedback, rendering the system more sensitive to 
variations in the delay time $\tau$.
The coherence of noise-induced oscillations measured by the correlation time exhibits
sharp resonances as a function of $\tau$, and can be strongly increased by 
optimal choices of $\tau$. 
Similarly, the peaks in the power spectral density are sharpened.
We provide analytical insight into the control mechanism by relating the
correlation times and mean frequencies of noise-induced breathing oscillations to the 
stability properties of the deterministic stationary current density filaments 
under the influence of the control loop. 
Moreover, we demonstrate that the use of multiple time delays enlarges the regime 
in which the deterministic dynamical properties of the system are 
not changed by delay-induced bifurcations.

\end{abstract}


\maketitle

\section{Introduction}
\label{sec:intro}

It is well known that random fluctuations can seriously affect charge transport in
semiconductors~\cite{BLA00}, which usually leads to deterioration of their
performance. However, recently the constructive role of noise in
semiconductor devices has been recognized. 
In particular, noise was shown to induce coherent radiation in semiconductor lasers
\cite{GIA00,SHE03,USH05} or to induce moving field domains in semiconductor
superlattices~\cite{HIZ06} whose regularity becomes optimum
at some non-zero value of the noise intensity. This phenomenon is known 
as {\em coherence resonance} \cite{HU93a,PIK97}. 
In double-barrier resonant tunneling (DBRT) diodes noise can generate 
spatially inhomogeneous current density patterns in form of breathing current
filaments~\cite{STE05}, however, their regularity decreases monotonically with increasing
noise intensity, and thus shows no coherence resonance. 

The control of the features of noise-induced dynamics is generally of great importance,
and has recently attracted a lot of attention in the field of nonlinear dynamic systems
\cite{SCH07}.
In~\cite{JAN03,BAL04} a method for manipulation of
essential features of noise-induced oscillations, like coherence and time scales, 
was proposed using a delayed feedback
scheme that was originally used to control chaos in purely deterministic
systems~\cite{PYR92}. 
This technique was demonstrated to be effective for
control of noise-induced oscillations in either simple (generic)
systems~\cite{SCH04b,POM05a,HAU06,PRA07,POT07,POT08} or more complex, spatially
extended systems\cite{HIZ05,BAL06,GAS08}. For the DBRT nanostructure 
it was shown that time-delayed feedback 
can either increase of decrease the regularity of noise-induced breathing filaments
and, moreover, can even lead to spatial homogenization of current density patterns
~\cite{STE05a}. For deterministic systems the original \emph{single} time
delayed feedback was extended by using multiples of the delay time $\tau$ weighted with a 
memory parameter \cite{SOC94}, which generally leads to larger control domains
and more efficient control \cite{BEC02,UNK03,SCH03a,DAH07}. In a simple stochastic
Van der Pol oscillator this has also been shown to yield drastically increased 
correlation times \cite{POM07}. 

In the present work we study the effect of \emph{multiple} time delayed 
feedback on the stochastic spatio-temporal pattern formation in the DBRT
model. Compared to the  \emph{single} time
delayed feedback in the same system~\cite{STE05a}, we find much sharper resonances
of the spectral and correlation properties in dependence upon the delay time.
For parameter values close to, but below, a Hopf bifurcation we show that 
in these sharp, pronounced resonances the temporal regularity is significantly increased 
and the power spectral width becomes much narrower. 
Moreover, we demonstrate that the use of multiple time delays enlarges the control 
parameter regime in which the original deterministic dynamical properties of the system do 
not change, i.e. the delay-induced bifurcations occur only at larger feedback strength.

The paper is organized as follows. In Section~II the DRBT model is
described, and the dynamical properties of the system for our chosen
parameters are discussed. Section~III is devoted to the effects of the
multiple time delayed feedback upon noise-induced dynamics, 
and in Section~IV we draw conclusions.

%%%%%%%%%%%%%%%%%%%%%%%%%%%%%%%%%%%%%%%%%%%%%%%%%%%%%%%%%%%%%%%%%%%%%%

\section{Model}
\label{sec:dbrt+noise}

In our study, we use a deterministic model for the DBRT suggested in \cite{SCH02} and add
two sources of random fluctuations as proposed in~\cite{STE05}.
Furthermore we use the time delayed feedback scheme
which was already applied to this system in~\cite{STE05a} and extend
it in order to take multiple time intervals into account in the
feedback loop:
\begin{equation}
  \label{eq:dyn_system}
  \begin{aligned}
    \frac{\partial a(x,t)}{\partial t} 
    &= f(a,u) + \frac{\partial}{\partial x} \left( D(a) \frac{\partial
        a}{\partial x} \right) + D_a\xi(x,t) \\
    \frac{\partial u(t)}{\partial t} 
    &= \frac{1}{\varepsilon } \left(U_0 - u - r J\right)  
    + D_u\eta(t)  + F(t) 
  \end{aligned}
\end{equation}
where all quantities are dimensionless. 
The dynamical variable $a(x,t)$ describes the
charge carrier density inside the quantum 
well, whereas $u(t)$ is the voltage drop across the device.
The spatial coordinate $x$ denotes the direction perpendicular to the
current flow, and $t$ is time.
In terms of nonlinear dynamics, in the deterministic
($D_a=D_u=0$) and uncontrolled ($K=0$) case, this is a
reaction-diffusion model of activator-inhibitor type, where $a$ is the
activator, and $u$ is the inhibitor \cite{SCH01}.

The net tunneling rate of the electrons through the two
energy barriers into and out of the quantum well is modelled by the nonlinear
function~\cite{SCH02}:
\begin{multline}
  \label{eq:f(a,u)}
    f(a,u) = j_{in}-j_{out}\\
j_{in}=\left[ 
      \frac{1}{2}+\frac{1}{\pi} \arctan \left(
        \frac{2}{\gamma}\left(x_0-\frac{u}{2}+\frac{d}{r_\text{B}}a\right)
        \right) \right] \\
       \times \left[
        \ln\left( 1+\exp\left(
        \eta_e-x_0+\frac{u}{2}-\frac{d}{r_\text{B}}a\right)
      \right) -a \right]\\
j_{out}= a,
\end{multline}
where $d$ is the effective thickness of the double-barrier structure, 
$r_\text{B}=(4\pi \epsilon \epsilon_0 \hbar^2)/(e^2m)$ is the
effective Bohr radius in the semiconductor material, $\epsilon$ and
$\epsilon_0$ are the relative and absolute permittivity of the material, and
$x_0$ and $\gamma$ describe the energy level and the broadening of the
electron states in the quantum well and $\eta_e$ is the dimensionless Fermi
level in the emitter, all in units of $k_\text{B}T$. Throughout the paper
we use values of $\gamma=6$, $d/r_\text{B}=2$, $\eta_e=28$ and $x_0=114$,
corresponding to typical device parameters at 4 K \cite{SCH02}.

The effective diffusion coefficient $D(a)$ resulting from the
inhomogeneous lateral redistribution of carriers and from the change in
the local potential due to the charge accumulated in the
quantum well by Poisson's equation is given by~\cite{CHE00}:
\begin{equation}
  \label{eq:D(a)}
  D(a) = a \left( \frac{d}{r_\text{B}} + \frac{1}{1-\exp(-a)}\right).
\end{equation}
It describes the diffusion
of the electrons within the quantum well perpendicular to the
current flow.
$J=\frac{1}{L}\int_0^L jdx$ gives
the total current through the device, where 
$j(a,u)=\frac{1}{2}\left(j_{in}+j_{out}\right)=\frac{1}{2}\left(f(a,u)+2a\right)$ is the 
local current density within the well.
The system's width is fixed at a value of $L=30$ and homogeneous Neumann boundary
conditions are used.

In Eq.~\eqref{eq:dyn_system} we use uncorrelated Gaussian white
noise sources $\xi(x,t)$ and $\eta(t)$ with noise intensities $D_a$ and $D_u$.
\begin{equation}
  \label{eq:noise}
  \begin{aligned}
  \langle\xi(x,t)\rangle=\langle\eta(t)\rangle &= 0 \qquad (x\in[0,L]), \\
  \langle\xi(x,t)\xi(x',t')\rangle &= \delta(x-x')\delta(t-t'), \\
  \langle\eta(t)\eta(t')\rangle &= \delta(t-t').
\end{aligned}
\end{equation}
Physically, $D_u$ can be realized by an external tunable noise voltage
source in parallel with the supply bias.
$D_a$ describes internal fluctuations of the local current density which
could be caused, e.\,g., by shot noise~\cite{BLA00}. 

The control force F(t) represents a control voltage which is constructed 
recursively from a time-delayed feedback loop with delay time $\tau$,
feedback strength $K \ge 0$, and memory parameter $R$, and can be written as
\begin{eqnarray}
F(t) &=& K(u(t-\tau)-u(t)) + R F(t-\tau)\\
     &=& K \sum_{n=0}^{\infty} R^n \left[u(t-(n+1)\tau) - u(t-n \tau) \right]. 
\end{eqnarray} 
 
The first Eq.~\eqref{eq:dyn_system} is the local balance
equation of the charge in the quantum well, and the second equation
represents Kirchhoff's law of the circuit in which the device is operated.
The control parameters are the external bias voltage $U_0$, the
dimensionless load resistance $r$, and the time-scale ratio
$\varepsilon=RC/\tau_a$, which is related to the load resistance $R$,
and the parallel capacitance $C$ of the attached circuit, normalized by
the tunneling time $\tau_a$.
A discussion of the various deterministic bifurcation 
scenarios can be found in~\cite{MEI00b,SCH02,UNK03}.

We fix $\varepsilon=6.2$ slightly below the Hopf bifurcation, which occurs at 
$\varepsilon_\text{Hopf} \approx 6.469$. In this regime we have two
fixed points: (i) a stable, spatially inhomogeneous fixed point
and (ii) a spatially homogeneous fixed point which is 
stable with respect to completely homogeneous perturbations but generally
unstable against spatially inhomogenous fluctuations (saddle-point).
Although the deterministic systems rests in the inhomogeneous steady state,
noise can induce irregular spatio-temporal oscillations of the current density
\cite{STE05}. In the following we shall study how
these noise-induced oscillations are influenced by the control force.

\section{Multiple time delayed feedback control}
\label{sec:control}

\begin{figure}[htbp]
  \centering
  \includegraphics[width=\linewidth]{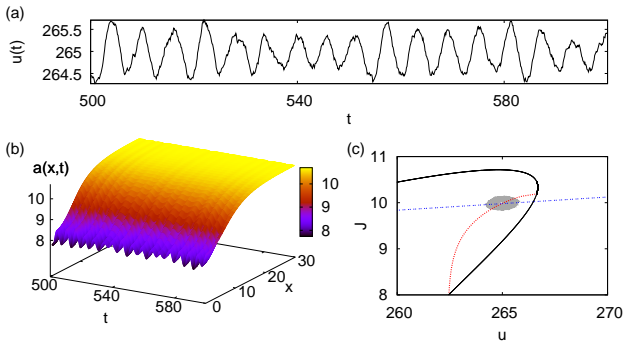}
  \caption{(color online) Stochastic spatio-temporal dynamics under
  multiple time-delayed feedback control. 
(a) Voltage time series $u(t)$ (in units of $0.35\,\text{mV}$), 
(b) charge carrier density $a(x,t)$ (in units of $10^{10}/\text{cm}^2$), 
(c) phase portrait of current $J$ (in units of $500\,\text{A}/\text{cm}^{2}$) 
vs. voltage $u$. Space $x$ and time $t$ are scaled in units of $100$ nm and $3.3$ ps, 
respectively, corresponding to typical device parameters at 4 K \cite{SCH02}.
Parameters are $U_0=-84.2895$, $r=-35$, $\epsilon=6.2$, $D_u=0.1$, $D_a=10^{-4}$, 
$K=0.1$, $\tau=6.3$, $R=0.5$.}
  \label{fig:space-time}
\end{figure}

\begin{figure}[htbp]
  \centering
  \includegraphics[width=\linewidth]{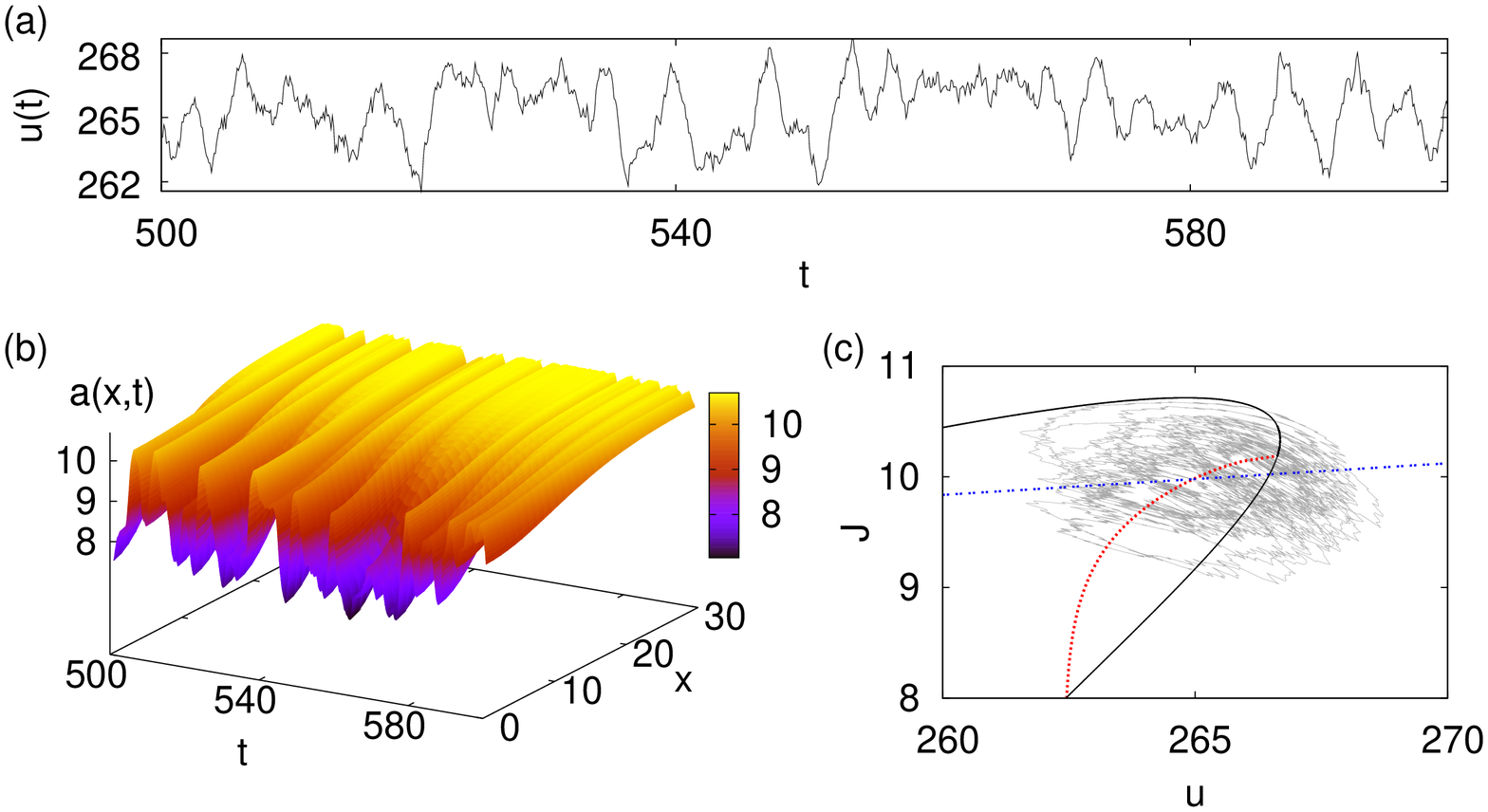}
  \caption{(color online) Same as Fig.~\ref{fig:space-time} for $D_u=1.0$.}
  \label{fig:space-time2}
\end{figure}

\begin{figure}[htbp]
  \centering
  \includegraphics[width=\linewidth]{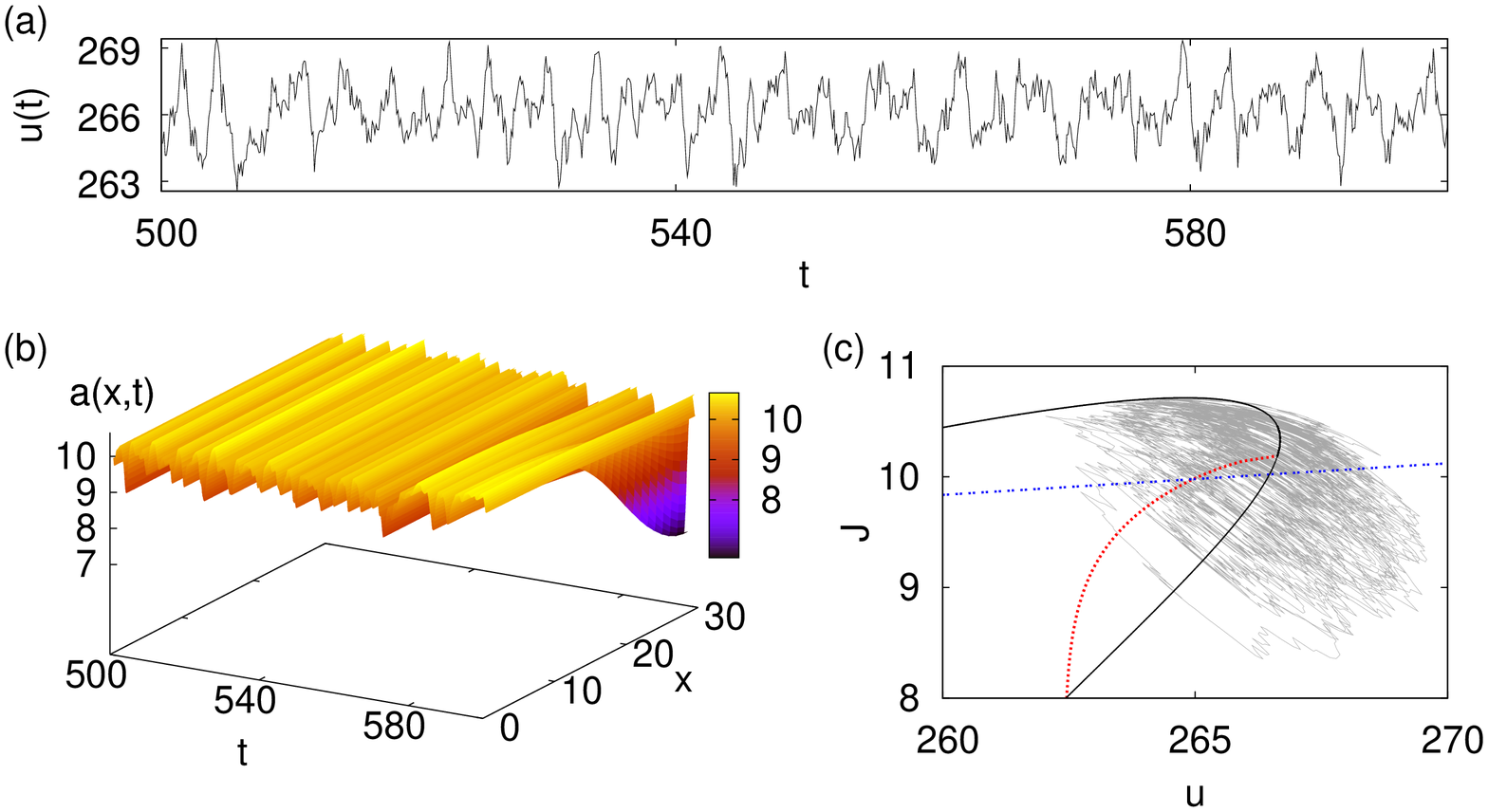}
  \caption{(color online) Same as Fig.~\ref{fig:space-time} for $D_u=2.0$.}
  \label{fig:space-time3}
\end{figure}

Fig.\ref{fig:space-time} shows simulations of the spatio-temporal
dynamics under the influence of noise and delayed feedback. The voltage time series (a),
the spatio-temporal charge density patterns (b), and the current-voltage projection of 
the infinite-dimensional phase space (c) are depicted.
Noise induces small \emph{spatially inhomogeneous} oscillations around the inhomogeneous
steady state ({\em breathing current filaments}). In the $J$-$u$ phase portrait (c), 
the spatially inhomogeneous steady state (fixed point)
is determined by the intersection of the load line (null isocline $\dot{u}=0$, blue 
dash-dotted) with the nullcline $\dot{a}=0$ for inhomogeneous $a(x,t)$ (red dotted).
The neighboring intersection of the load line with the nullcline
$\dot{a}=0$ for homogeneous $a$ (black solid curve) defines the second, 
spatially homogeneous fixed point which is a saddle. 
With increasing noise intensity (Fig.\ref{fig:space-time2})
the oscillation amplitude becomes larger, the oscillations become more irregular, and
finally, at even larger noise, the oscillations are more spatially homogeneous, 
i.e., in the phase space they are more centered around the homogeneous fixed point 
(Fig.\ref{fig:space-time3}).

We shall now investigate how the regularity and the time-scales of these noise-induced
oscillations depend upon the feedback parameters $K$, $R$, $\tau$.

\subsection{Linear stability analysis}
\label{sec:char_equation}

To this purpose we first examine the stability properties of the inhomogeneous fixed 
point ($a_0(x), u_0$) under the influence of the control force. We perform a linearization 
of the original continuous system~(\ref{eq:dyn_system}) (with $D_u = D_a = 0$) around the 
inhomogeneous fixed point along the same lines as in \cite{ALE98, STE05a}. We use an 
exponential ansatz for the deviations from the fixed point $\delta a(x,t) \equiv 
a(x,t) - a_0(x) = e^{\Lambda t} \tilde a(x)$ and $\delta u(t) \equiv u(t) - u_0 = 
e^{\Lambda t} \tilde u$. 
The resulting coupled eigenvalue problem reads 
\begin{align}
 \Lambda \tilde{a}(x) &= \hat{H}\tilde{a}(x)+f_u(x)\tilde{u}, \label{eq:a} \\
\Lambda \tilde{u} &= 
    -\frac{r}{\varepsilon L} \int_0^L j_a(x)\tilde{a}(x)dx
    -\frac{1+rJ_u}{\varepsilon} \tilde{u} \notag \\
&\quad \,-K\frac{1 - \operatorname{e}^{-\Lambda
          \tau}}{1 - R \operatorname{e}^{- \Lambda \tau}} \tilde{u}, \label{eq:u}
\end{align}
where we have introduced a self-adjoint linear operator $\hat H$. 
%\begin{equation}
%    \begin{aligned}[t]
%      \hat H \equiv \biggl.\frac{\partial f}{\partial
%        a}\biggr|_{a_0,u_0} & + \biggl.\frac{\partial b}{\partial
%        a}\biggr|_{a_0} + \biggl.\frac{\partial b}{\partial
%        a_x}\biggr|_{a_0}
%      \frac{\partial}{\partial x} \\
%      &+ \biggl.\frac{\partial b}{\partial a_{xx}}\biggr|_{a_0}
%      \frac{\partial^2}{\partial x^2},
%  \end{aligned}
%\end{equation}
Its eigenfunctions $\Psi_i$ and eigenvalues $\lambda_i$ correspond to the voltage-clamped
system,  $\delta u = 0$. 
Furthermore,
\begin{align}  
  \begin{gathered}[t]
    f_u \equiv \biggl.\frac{\partial f}{\partial u}\biggr|_{a_0,u_0}, \quad
    j_a \equiv \biggl.\frac{\partial j}{\partial a}\biggr|_{a_0,u_0}, \\
    J_u =
    \frac{1}{L}\int_0^L \biggl.\frac{\partial j}{\partial
      u}\biggr|_{a_0,u_0} dx.
\end{gathered}
\end{align}
Due to the global constraint, Eq.(\ref{eq:u}) mixes the 
eigenmodes $\Psi_i$ and both equations have to be solved simultaneously. 
An expansion of the eigenmodes $\tilde a$ of the full system in terms of the eigenmodes 
$\Psi_i$ of the voltage-clamped system, keeping only the dominant eigenmode $\Psi_0$ with
eigenvalue $\lambda_0>0$, 
leads to a characteristic equation for the eigenvalues $\Lambda$ 
 \begin{multline}
  \label{eq:char_eq1}
  \Lambda^2 + \left(\frac{1+rJ_u}{\varepsilon}-\lambda_0\right) \Lambda
  \\+ (\Lambda -\lambda_0) K\frac{ 1 -\operatorname{e}^{-\Lambda \tau}}{1 - R \operatorname{e}^{- \Lambda \tau}}
  -\frac{\lambda_0}{\varepsilon}(1+r\sigma_d) =0,
\end{multline}
where the static differential conductance at the inhomogeneous fixed point $\sigma_d \equiv \left.\frac{dJ}{du}\right|_{a_0,u_0}$
has been introduced.
In~\cite{STE05a} a more detailed derivation of the characteristic equation is given
for the special case $R=0$. The extension to the case $R \neq 0$ is straightforward.

Without control, $K=0$, eq.~\eqref{eq:char_eq1} reduces to a
characteristic polynomial of second order, which gives the well-known
conditions for stability of a filament~\cite{ALE98}
\begin{equation}
  \label{eq:stability}
  \begin{aligned}
    A &\equiv \frac{1+rJ_u}{\varepsilon}-\lambda_0>0, \\
    C &\equiv -\frac{\lambda_0}{\varepsilon}(1+r\sigma_d)>0.
  \end{aligned}
\end{equation}
and a Hopf bifurcation occurs on the two-dimensional center manifold
if $A=0$. 

With control, eq.~\eqref{eq:char_eq1} can be expressed as
\begin{equation}
  \label{eq:char_eq}
 \Lambda^2 + A \Lambda + (\Lambda -B)K\frac{
      1 - \operatorname{e}^{-\Lambda \tau}}{1 - R \operatorname{e}^{-\Lambda \tau}} +C =0
\end{equation}
with $B\equiv\lambda_0>0$. 
The parameters $A$, $B$, $C$ can be calculated
directly from ~\eqref{eq:stability} \cite{STE05a}, yielding
$A=0.0447$, $B=1.0281$ and $C=1.1458$.

Using Eq.~(\ref{eq:char_eq}) we can calculate the domains of stability in the $\tau$-$K$ 
plane numerically for selected values of the memory parameter $R$. 
In order to find the curves containing the boundaries of stability of the inhomogeneous 
fixed point as a subset, 
we set $\Lambda=p+i q$ with $p=0$ 
and separate Eq.~\eqref{eq:char_eq} into real and imaginary parts:
\begin{eqnarray}
 \left[BK -R \left( C-q^2\right) \right] \cos(q \tau) - q \left( AR + K\right) \sin(q \tau)& \notag \\ 
= BK + \left( q^2 -C\right) &\label{four}
\end{eqnarray}
\begin{eqnarray}
 q \left( AR + K\right) \cos(q \tau) + \left[ BK +R \left( q^2 - C\right)\right] \sin(q \tau)& \notag \\
= q \left(A + K \right)& \label{three}
\end{eqnarray}

Using Eq. \eqref{four} and Eq. \eqref{three} the boundary of stability can be 
obtained from the set of parametric functions $K(q)$ and $\tau(q)$ using  
$q=Im(\Lambda)$ as the curve parameter. 
\begin{widetext}
\begin{eqnarray}
K(q)&=&\frac{\left(A^2 q^2 + (C - q^2)^2 \right)(1+R)}{2 B C -2 (A + B) q^2} \notag  \\
\tau(q)&=& \frac{1}{q} \left( \arcsin \left( \frac{q(AB+C-q^2)(1-R)K}{(A^2 q^2+(C-q^2)^2)R^2+2 (-BC+(A+B)q^2)R K+(B^2+q^2)K^2} \right) + 2 \pi N  \right) \label{keq}
\end{eqnarray}
\end{widetext}

\begin{figure}[htbp]
  \centering
  \includegraphics[width=\linewidth]{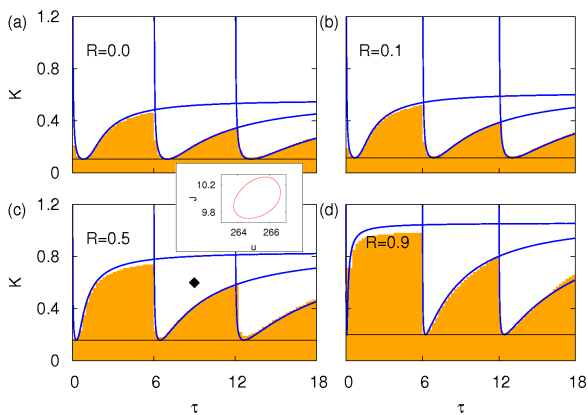}
  \caption{(color online) (a)-(d) Stability domains of the inhomogeneous fixed point in 
the $\tau$-K plane of the deterministic system \eqref{eq:dyn_system} ($D_u=D_a=0$), 
for selected values of the memory parameter $R$. 
Blue lines:
Solutions of \eqref{eq:char_eq} with $Re(\Lambda)=0$ calculcated from  Eq. \eqref{keq}. 
Orange (shaded) region: regime of stability of the fixed point obtained from the 
numerical solution of Eq. \eqref{eq:dyn_system}. 
Black horizontal line: upper bound for K where the fixed point is stable 
for all values of $\tau$, calculated from  Eq.\eqref{threshold}. 
The black diamond in panel (c) marks the parameter values for which a $J-u$ phase portrait 
of the delay-induced limit cycle is shown in the inset.} 
  \label{fig:ktau}
\end{figure}

Fig.~\ref{fig:ktau} shows these curves, Eq. \eqref{keq}, as blue lines.
The boundaries of stability, where a delay-induced Hopf bifurcation of deterministic
breathing oscillations occurs, are a subset of these curves, because the fixed point
may already be unstable when a complex eigenvalue crosses the imaginary axis, due to 
other unstable eigenvalue branches. The boundaries are in good agreement with  
the domain of stability obtained from dynamical simulations of the nonlinear system 
equations \eqref{eq:dyn_system} ($D_u=D_a=0$), shown as orange (dark shaded) areas.
The inset of panel (c) shows the delay-induced limit cycle in the $J-u$ phase space for 
parameters outside the stability domain of the inhomogeneous fixed point.
The stability domains increase significantly with increasing memory parameter $R$ from
(a) to (d). The modulation of their boundaries in dependence on $\tau$ results from the 
cross-over of differerent eigenvalue branches, which is a typical feature of delay 
differential equations.

From Eq. \eqref{keq} it is possible to calculate an upper bound $K_c$ of $K$ for 
which the stability properties of the uncontrolled deterministic system remain unchanged 
over the whole $\tau$ interval, meaning that no delay-induced Hopf bifurcation occurs.
\begin{eqnarray}\label{threshold}
K_c &=& \frac{A^2 (A+B) \left( BC - G \right) + \left( AC + G \right)^2}{2 (A+B)^2 G} (1+R) \notag \\
&\approx& 0.1059 (1+R) \notag \\
G:&=&\sqrt{A^2 C (B (A+ B) +C)}
\end{eqnarray}
Fig.~\ref{fig:ktau} shows this upper bound plotted as black horizontal line.

\subsection{Correlation times}
\label{sec:numerics}

\begin{figure}[htbp]
  \centering
  \includegraphics[width=\linewidth, height=5.6cm]{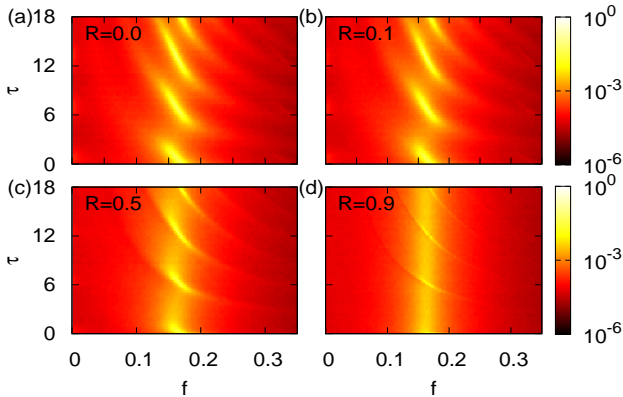}
  \caption{(color online) (a)-(d) Power spectral density $S_{uu}(f)$ of the dynamical variable 
$u$ in dependence of the frequency $f$ and the delay time $\tau$ for selected 
values of $R$ ($K=0.1$, $\epsilon=6.2$, $D_u=0.1$, $D_a=10^{-4}$).} 
  \label{fig:pow_spec_numeric}
\end{figure}
To quantify the temporal regularity of the noise-induced oscillations, 
we evaluate
the correlation time \cite{STR63} calculated from the voltage signal,
\begin{equation}
  \label{eq:t_cor}
  t_\text{cor} \equiv \frac{1}{\sigma^2} \int_0^\infty \left|\Psi(s)\right|ds,
\end{equation}
where $\Psi(s) \equiv \bigl\langle\left(u(t)-\langle u\rangle\right)
  \left(u(t+s)-\langle u\rangle\right)\bigr\rangle_t$ is the autocorrelation
  function of the variable $u(t)$ and $\sigma^2=\Psi(0)$ its variance.

In order to investigate the influence of multiple time delayed feedback control 
in the DBRT, we systematically study the 
dependence of the correlation time from Eq.~(\ref{eq:t_cor}) upon the control force 
parameters $\tau$, $K$, and $R$.

\begin{figure}[htbp]
  \centering
  \includegraphics[width=\linewidth]{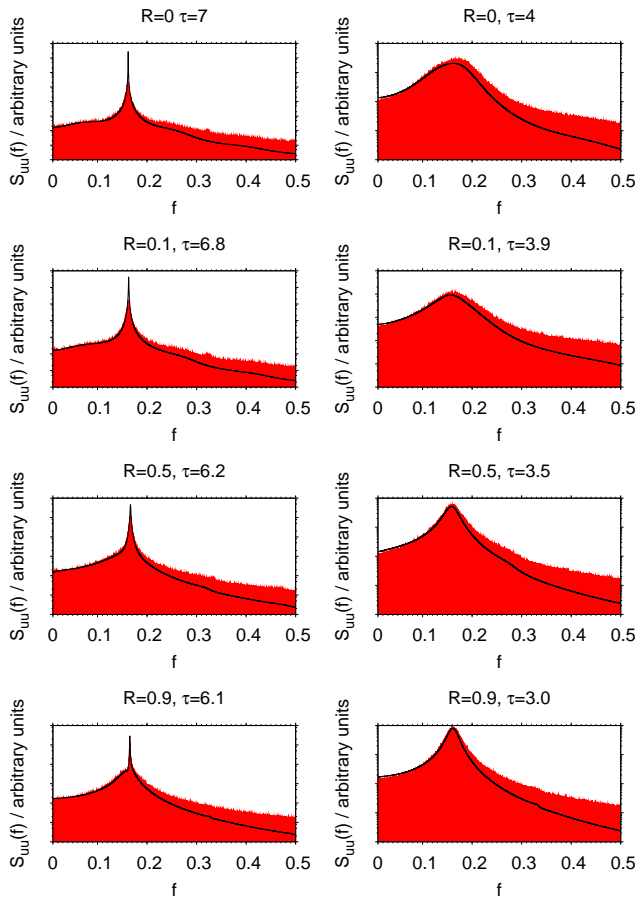}
  \caption{(color online) Power spectral density $S_{uu}(f)$ of the dynamical variable 
$u$ in dependence of the frequency $f$ for various delay times $\tau$ 
and memory parameters $R$  
($K=0.1$, $\epsilon=6.2$, $D_u=0.1$, $D_a=10^{-4}$).} 
  \label{fig:pow_spec_analytic}
\end{figure}

In Fig.~\ref{fig:pow_spec_numeric} the Fourier power spectral density $S_{uu}$ 
obtained from the time series $u(t)$ 
is shown in dependence of the delay time $\tau$ for different values of 
the memory parameter $R$. The shape of the spectra $S_{uu}(2\pi f)$ alternates between broad and 
sharply peaked with varying $\tau$. This shows up more clearly in the corresponding sections
at fixed $\tau$ depicted in Fig.~\ref{fig:pow_spec_analytic}. 
A very good analytic approximation of the power spectral density can be obtained by a
straightforward extension of the argument in \cite{STE05a} to multiple time-delayed 
feedback:

\begin{widetext}
\begin{align}
S_{uu}(\omega) =  &\frac{{D^\prime}^2}{2 \pi} \left[  \left(  -\omega^2 + BK \frac{(\cos{(\omega \tau)} -1 ) (R+1)}{1 + R^2 -2R\cos{(\omega \tau)}} - \frac{\omega K (1 -R ) \sin{(\omega \tau)}}{1+ R^2 -2R\cos{(\omega \tau)}} + C \right)^2 \right. \notag \\ 
& + \left. \left( -A \omega + \omega K \frac{(\cos{(\omega \tau)} -1) (R + 1)}{1 + R^2 - 2 R \cos{(\omega \tau)}} + B K \frac{(1 - R ) \sin{(\omega \tau)}}{1 + R^2 - 2R \cos{(\omega \tau)}}\right)^2  \right] ^{-1
}.
\label{eq:pow_spec_density}
\end{align}
\end{widetext}
which is shown as black curves in Fig.~\ref{fig:pow_spec_analytic}.

\begin{figure}[htbp]
  \centering
  \includegraphics[width=\linewidth]{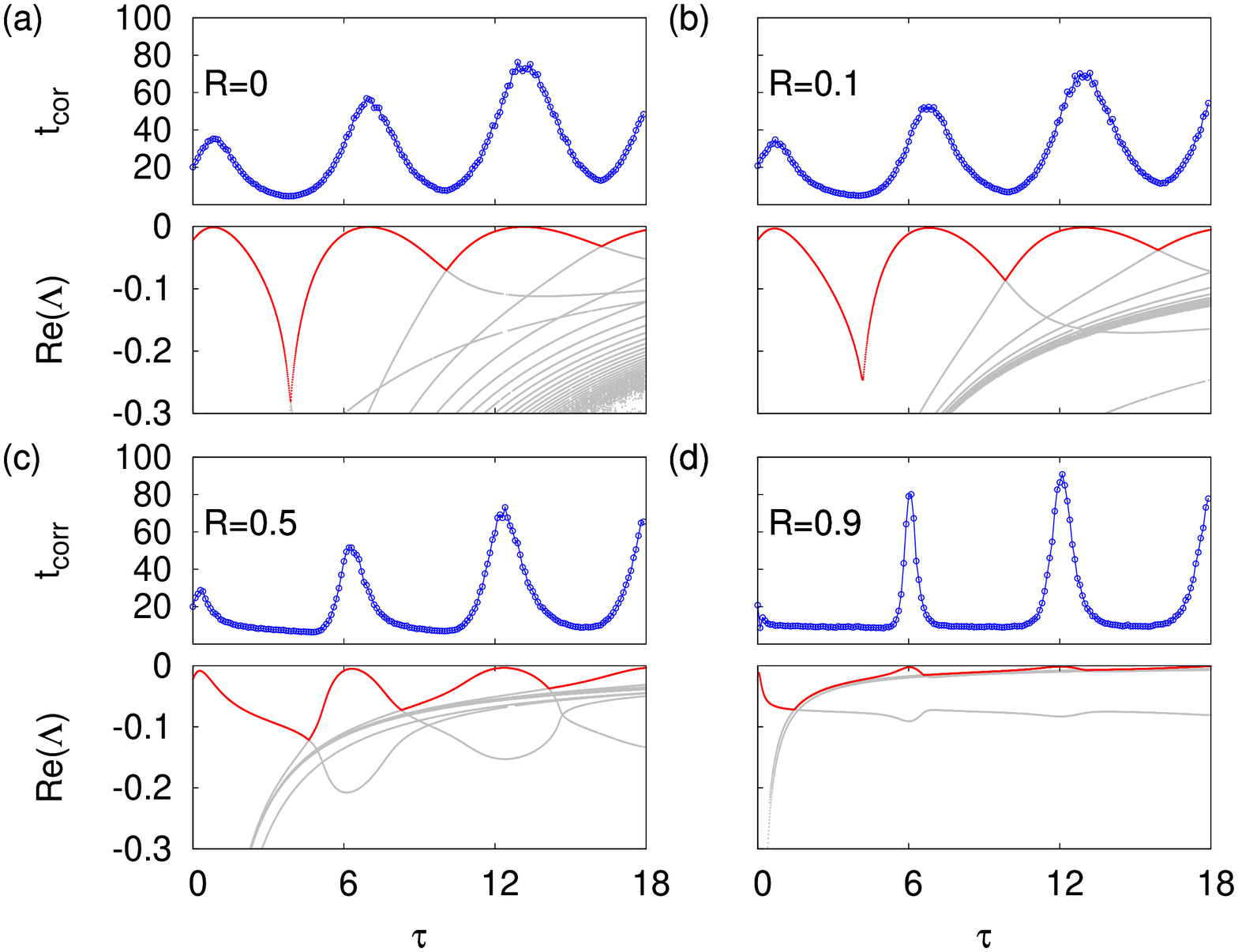}
  \caption{(color online) Correlation times $t_{cor}$ (upper panels) and deterministic stability of the 
inhomogeneous fixed point, $Re(\Lambda)$ (lower panels), in dependence of the delay 
time $\tau$ for different values of the memory parameter $R$ (a)-(d) and fixed $K=0.1$. 
The red (dark) curves in the lower panels mark the leading eigenvalue, which governs the 
overall stability of the fixed point. Parameters: $\epsilon=6.2$, 
$D_u=0.1$ and $D_a=10^{-4}$ (in the panels showing $t_{cor}$).} 
  \label{fig:tc_tau}
\end{figure}

At certain resonant values
of $\tau$ (left column) the spectral peaks become extremely sharp.
With increasing memory parameter $R$ the 
broad spectra prevail over larger intervals of $\tau$, whereas the regime of sharply 
peaked spectra becomes smaller. Thus multiple time feedback control exhibits much
more pronounced resonant features both in the frequency domain and in the delay time. 
Since a sharply peaked spectrum gives rise to long correlation times (which are, 
in the linear regime, proportional to the inverse spectral width) we expect the 
domains of strong correlation to shrink with increasing memory parameter and the 
domains of low correlation to increase. Extracting from the Fourier power spectral 
density the autocorrelation function $\Psi(s)=\int_{-\infty}^\infty S_{uu}(f) e^{2 \pi 
i f s } \, df$ and using Eq.~(\ref{eq:t_cor}) we obtain the correlation time $t_{cor}$ 
in dependence of $\tau$. This is shown in
Fig.~\ref{fig:tc_tau} for different values of the memory parameter $R$. The feedback 
strength is kept at a constant value of $K=0.1$, where the system is  below the Hopf 
bifurcation for all values of $\tau$ and $R$.

For small memory parameter $R$ the correlation times alternate between high and low 
values with growing $\tau$. For higher memory parameters $R$ the peaks in correlation 
time indeed become narrower and sharpen up, and the domains of low correlation time 
increase.
The stability of the inhomogeneous fixed point reveals a relation between properties of 
the 
controlled deterministic system and the noise-induced dynamics: maximum regularity of 
noise-induced oscillations is attained when the deterministic fixed point is least stable. 
This feature is maintained for all values of the memory parameter $R$.
In the case of small $R$ the crossover of the real part of eigenvalue branches also 
determines the 
location of the minima in correlation time. In that case two eigenmodes with the same 
stability (real part) but different frequencies are present in the system, resulting in 
rather irregular noise-induced dynamics.  For large memory parameters the broad domains of 
low correlation display many eigenmodes that are not well separated (stability-wise) 
causing irregular mixed dynamics. 
\begin{figure}[htbp]
  \centering
  \includegraphics[width=\linewidth, height=5.6cm]{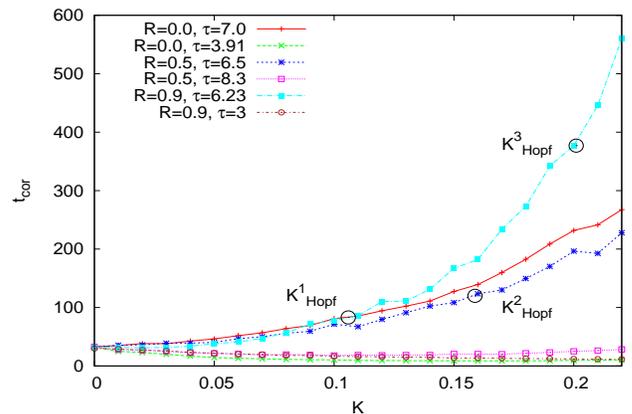}
  \caption{(color online) Correlation times $t_{cor}$ in dependence of $K$ for different 
values of the memory parameter $R$ and optimal and non-optimal $\tau$ 
($\epsilon=6.2$, $D_u=0.1$, $D_a=10^{-4}$). $K^1_{Hopf}$, $K^2_{Hopf}$ and $K^3_{Hopf}$ 
mark the values of K at which the Hopf bifurcation occurs for R=0, R=0.5, and R=0.9, 
respectively ($\epsilon=6.2$, $D_u=0.1$, $D_a=10^{-4}$).} 
  \label{fig:tc_k}
\end{figure}

The regularity of the noise-induced oscillations in dependence of the 
control force strength $K$ is visualized in Fig.~\ref{fig:tc_k}. The correlation time
vs $K$ is shown  
for different values of $R$ and $\tau$. Depending upon the chosen value of $\tau$ the 
correlation time increases or decreases with growing $K$. An optimal value of $\tau$ 
leads to more regular oscillations whereas a non-optimal value of $\tau$ results in more 
irregular oscillations. In the case of an optimally chosen $\tau$ and a value of $K>0.1$ 
the curves split for larger $K$, and the one with largest $R$ (blue full squares)
attains the highest correlation times, clearly above the curves  
for $R=0$ (red plus) and $R=0.5$ (violet asterisks). 
Comparing the values of $t_{cor}$ at the threshold of delay-induced Hopf bifurcation
of the fixed point, marked by $K^i_{Hopf}$, the order of 
increasing correlation time is from small to large $R$.
\begin{figure}[htbp]
  \centering
  \includegraphics[width=\linewidth]{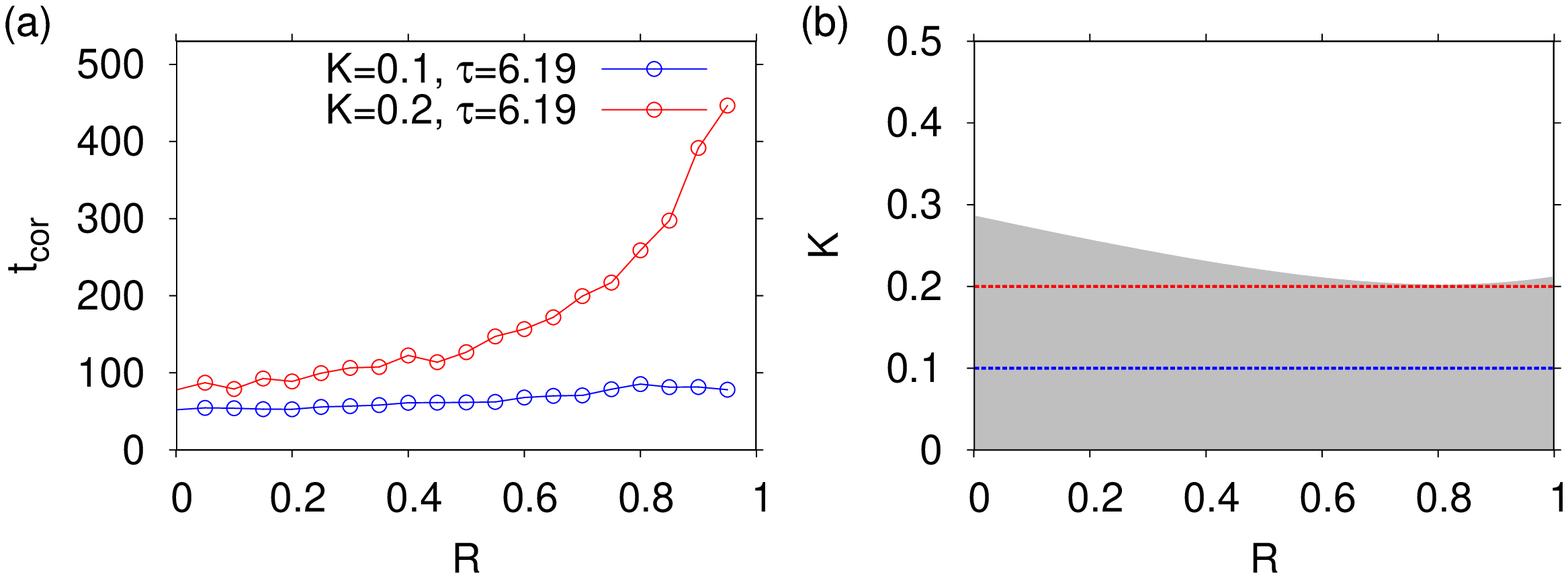}
  \caption{(color online)(a): Correlation times $t_{cor}$ in dependence of the memory 
parameter $R$ vor different values of the coupling strength $K$ and $\tau=6.19$. 
(b): Domain of stability of the inhomogeneous fixed point (green shaded) in the ($K,R$) 
plane for $\tau=6.19$. The horizontal lines mark the values of $K$ chosen in (a) 
($\epsilon=6.2$, $D_u=0.1$, 
$D_a=10^{-4}$). } 
  \label{fig:tc_r}
\end{figure}

The dependence of the correlation time on the memory parameter $R$ is shown in 
Fig.~\ref{fig:tc_r}(a) at a fixed value of $\tau=6.19$ and values of $K=0.1$ and $0.2$. 
Initially starting at a higher value of $t_{cor}$ the correlation time increases more 
rapidly and reaches higher values in the case of larger K (K=0.2). 
Figure~\ref{fig:tc_r}(b) shows the domain of stability in the $K$-$R$ plane. The colored 
(dark) region is the domain where the inhomogeneous fixed point is stable under the 
influence of the feedback loop without noise. The two lines mark the chosen values of 
$K$ used in Fig.~\ref{fig:tc_r}(a). Again, the correlation time is maximum if the 
fixed point is closest to its boundary of stability.

\subsection{Mean period}
\begin{figure}[htbp]
  \centering
  \includegraphics[width=\linewidth]{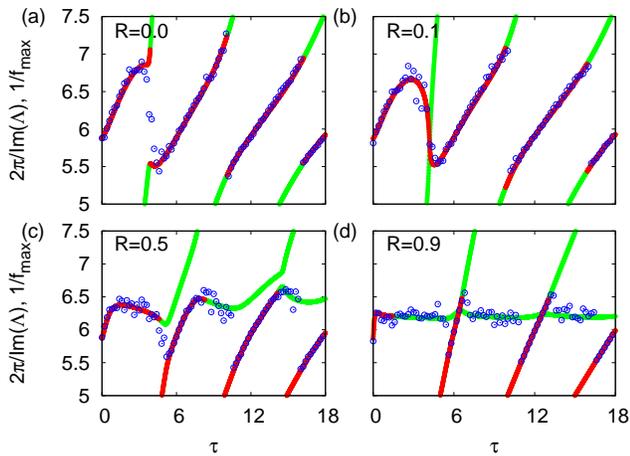}
  \caption{(color online) Main period $T_0=1/f_{max}$ of the noise induced oscillations 
obtained from the 
main peak in the power spectral density (blue open circles). Red (solid) curve: 
Period $T=2 \pi / Im(\Lambda_0)$ of the leading eigenvalue. Green (grey) curves: Period 
$T=2\pi/Im(\Lambda_i)$ ($K=0.1$, $\epsilon=6.2$, $D_u=0.1$, $D_a=10^{-4}$).} 
  \label{fig:timescales}
\end{figure}

Here we will establish a relation between the linear modes of 
the inhomogeneous fixed point and the time scales of the noise-induced oscillations in 
its vicinity. The stability and angular frequency of the eigenmodes are given by the real 
and imaginary parts of the eigenvalues $\Lambda_i$ of Eq.~(\ref{eq:char_eq}), 
$Re(\Lambda_i)$ and $Im(\Lambda_i)$, respectively.
Fig.~(\ref{fig:timescales}) shows a plot of the main periods of the noise-induced 
oscillations obtained from the main peak in the Fourier power spectral density (blue open 
circles) compared to the periods calculated from the imaginary part $Im(\Lambda_i)$ of 
the eigenvalues of the deterministic fixed point using Eq.~(\ref{eq:char_eq}). The red 
(solid) curve shows the periods $T_0=2 \pi/Im(\Lambda_0)$ of the leading eigenvalue. For 
small memory parameter $R$ the period of the oscillations closely follows the eigenperiod 
corresponding to the leading eigenvalue over the whole $\tau$ interval, whereas for 
larger $R$ this feature is only maintained in a narrow domain where the leading 
eigenvalue has a real part close to zero and is clearly separated from all other 
eigenmodes (compare Fig.~\ref{fig:tc_tau}). 

\section{Conclusion}
\label{sec:conclusion}
In conclusion, we have shown that multiple time-delayed feedback control 
leads to more pronounced resonant features of noise-induced spatio-temporal
current oscillations in a semiconductor nanostructure compared to single time feedback.
The regularity of noise-induced oscillations measured by the correlation time exhibits
sharp resonances as a function of the delay time $\tau$, and can be strongly increased by 
control with optimal choices of $\tau$, whereas it decreases in a broad range of 
non-optimal values of $\tau$. Thus the system is more sensitive to variations in $\tau$.
Similarly, the peaks in the power spectral density are sharper and exhibit stronger
resonances in dependence on $\tau$ for multiple time feedback, whereas their position, i.e., 
the main period of the oscillations, is less sensitive to variations in $\tau$ in wider 
intervals.

The regularity and time scales of noise-induced breathing oscillations are related to the 
stability properties of the deterministic stationary filamentary current pattern 
(fixed point) 
under the influence of the control loop. Maximum regularity is attained if the fixed point
is least stable in the deterministic case. In the domains of high temporal correlation the 
period of the noise-induced oscillations corresponds to the eigenperiod of the leading 
eigenvalue in the deterministic system. 

Furthermore, we have shown that 
using multiple time delayed feedback control compared to single time-delayed feedback 
leads to larger regimes of stability of the stationary filamentary current pattern 
in the deterministic system, and delay-induced bifurcations occur only at larger values
of the control amplitude $K$. 

\begin{acknowledgments}
This work was supported by DFG in the framework of Sfb 555. We are grateful to Grischa 
Stegemann for helpful discussion.
\end{acknowledgments}

%\bibliographystyle{/home/agschoell/latex-local/bst/prsty-fullauthor}
%\bibliography{/home/agschoell/latex-local/bib/ref}

\begin{thebibliography}{10}

\bibitem{BLA00}
Y.~M. Blanter and M. B{\"u}ttiker, Phys.~Rep. {\bf 336},  1  (2000).

\bibitem{GIA00}
G. Giacomelli, M. Giudici, S. Balle, and J.~R. Tredicce, Phys.~Rev.~Lett. {\bf
  84},  3298  (2000).

\bibitem{SHE03}
V.~V. Sherstnev, A. Krier, A.~G. Balanov, N.~B. Janson, A.~N. Silchenko, and
  P.~V.~E. McClintock, Fluct. Noise Lett. {\bf 3},  91  (2003).

\bibitem{USH05}
O.~V. Ushakov, H.~J. W{\"u}nsche, F. Henneberger, I.~A. Khovanov, L.
  Schimansky-Geier, and M.~A. Zaks, Phys.~Rev.~Lett. {\bf 95},  123903  (2005).

\bibitem{HIZ06}
J. Hizanidis, A.~G. Balanov, A. Amann, and E. Sch{\"o}ll, Phys.~Rev.~Lett. {\bf
  96},  244104  (2006).

\bibitem{HU93a}
G. Hu, T. Ditzinger, C.~Z. Ning, and H. Haken, Phys.~Rev.~Lett. {\bf 71},  807
  (1993).

\bibitem{PIK97}
A. Pikovsky and J. Kurths, Phys.~Rev.~Lett. {\bf 78},  775  (1997).

\bibitem{STE05}
G. Stegemann, A.~G. Balanov, and E. Sch{\"o}ll, Phys.~Rev.~E {\bf 71},  016221
  (2005).

\bibitem{SCH07}
{\em Handbook of Chaos Control}, edited by E. Sch{\"o}ll and H.~G. Schuster
  (Wiley-VCH, Weinheim, 2008), second completely revised and enlarged edition.

\bibitem{JAN03}
N.~B. Janson, A.~G. Balanov, and E. Sch{\"o}ll, Phys.~Rev.~Lett. {\bf 93},
  010601  (2004).

\bibitem{BAL04}
A.~G. Balanov, N.~B. Janson, and E. Sch{\"o}ll, Physica~D {\bf 199},  1
  (2004).

\bibitem{PYR92}
K. Pyragas, Phys.~Lett.~A {\bf 170},  421  (1992).

\bibitem{SCH04b}
E. Sch{\"o}ll, A.~G. Balanov, N.~B. Janson, and A. Neiman, Stoch.~Dyn. {\bf 5},
   281  (2005).

\bibitem{POM05a}
J. Pomplun, A. Amann, and E. Sch{\"o}ll, Europhys.~Lett. {\bf 71},  366
  (2005).

\bibitem{HAU06}
B. Hauschildt, N.~B. Janson, A.~G. Balanov, and E. Sch{\"o}ll, Phys.~Rev.~E
  {\bf 74},  051906  (2006).

\bibitem{PRA07}
T. Prager, H.~P. Lerch, L. Schimansky-Geier, and E. Sch{\"o}ll, J. Phys. A {\bf
  40},  11045  (2007).

\bibitem{POT07}
A. Pototsky and N.~B. Janson, Phys.~Rev.~E {\bf 76},  056208  (2007).

\bibitem{POT08}
A. Pototsky and N.~B. Janson, Phys.~Rev.~E {\bf 77},  031113  (2008).

\bibitem{HIZ05}
J. Hizanidis, A.~G. Balanov, A. Amann, and E. Sch{\"o}ll, Int.~J.~Bifur.~Chaos
  {\bf 16},  1701  (2006).

\bibitem{BAL06}
A.~G. Balanov, V. Beato, N.~B. Janson, H. Engel, and E. Sch{\"o}ll,
  Phys.~Rev.~E {\bf 74},  016214  (2006).

\bibitem{GAS08}
M. Gassel, E. Glatt, and F. Kaiser,
  Phys.~Rev.~E {\bf 77},  066220  (2008).

\bibitem{STE05a}
G. Stegemann, A.~G. Balanov, and E. Sch{\"o}ll, Phys.~Rev.~E {\bf 73},  016203
  (2006).

\bibitem{SOC94}
J.~E.~S. Socolar, D.~W. Sukow, and D.~J. Gauthier, Phys.~Rev.~E {\bf 50},  3245
   (1994).

\bibitem{BEC02}
O. Beck, A. Amann, E. Sch{\"o}ll, J.~E.~S. Socolar, and W. Just, Phys.~Rev.~E
  {\bf 66},  016213  (2002).

\bibitem{UNK03}
J. Unkelbach, A. Amann, W. Just, and E. Sch{\"o}ll, Phys.~Rev.~E {\bf 68},
  026204  (2003).

\bibitem{SCH03a}
J. Schlesner, A. Amann, N.~B. Janson, W. Just, and E. Sch{\"o}ll, Phys.~Rev.~E
  {\bf 68},  066208  (2003).

\bibitem{DAH07}
T. Dahms, P. H{\"o}vel, and E. Sch{\"o}ll, Phys.~Rev.~E {\bf 76},  056201
  (2007).

\bibitem{POM07}
J. Pomplun, A.~G. Balanov, and E. Sch{\"o}ll, Phys.~Rev.~E {\bf 75},  040101(R)
   (2007).

\bibitem{SCH02}
E. Sch{\"o}ll, A. Amann, M. Rudolf, and J. Unkelbach, Physica~B {\bf 314},  113
   (2002).

\bibitem{SCH01}
E.~Sch{\"o}ll: {\em Nonlinear spatio-temporal dynamics and chaos in
  semiconductors\/} (Cambridge University Press, Cambridge, 2001)

\bibitem{CHE00}
V. Cheianov, P. Rodin, and E. Sch{\"o}ll, Phys.~Rev.~B {\bf 62},  9966  (2000).

\bibitem{MEI00b}
M.~Meixner, P.~Rodin, E.~Sch{\"o}ll, and A.~Wacker, Eur.~Phys.~J.~B {\bf 13}, 157
  (2000).

\bibitem{ALE98}
A. Alekseev, S. Bose, P. Rodin, and E. Sch{\"o}ll, Phys.~Rev.~E {\bf 57},  2640
   (1998).

\bibitem{STR63}
R.~L. Stratonovich, {\em Topics in the Theory of Random Noise} (Gordon and
  Breach, New York, 1963), Vol.~1.

\end{thebibliography}

\end{document}